\documentclass[10pt,twocolumn]{IEEEtran}
\usepackage[T1]{fontenc}
\usepackage{amsthm}
\usepackage{amssymb}

\usepackage{epsfig}
\usepackage{extarrows}
\usepackage{graphicx}
 \usepackage{subfigure}
\usepackage{lipsum}
\usepackage{amsmath}
\usepackage{color}
\newcommand{\beq}{\begin{equation}}
\newcommand{\eeq}{\end{equation}}
\newcommand{\bea}{\begin{array}}
\newcommand{\ena}{\end{array}}
\usepackage{epstopdf}
\newtheorem{my_theorem}{Theorem}

\newtheorem*{other_theorems}{Theorem}
\newtheorem{my_lemma}{Lemma}

\newtheorem{definition}{Definition}

\begin{document}
\title{Distributed Energy Efficient Channel Allocation}

\author{
}

\author{Oshri Naparstek,~\IEEEmembership{Member,~IEEE}, S.~M.~ Zafaruddin,~\IEEEmembership{Member,~IEEE,}
	Amir Leshem,~\IEEEmembership{Senior Member,~IEEE}, 
Eduard  Jorswieck,~\IEEEmembership{Senior Member,~IEEE}

\thanks{Oshri Naparstek was a Ph.D. student with Faculty of Engineering, Bar-Ilan university, Ramat-Gan, 52900, Israel  (email: oshri8@gmail.com). S. M. Zafaruddin and Amir Leshem  are with  Faculty of Engineering, Bar-Ilan university, Ramat-Gan, 52900, Israel (email: smzafar@biu.ac.il, leshema@biu.ac.il). Eduard  Jorswieck is with the Faculty of Electrical and Computer Engineering,
	Communications Laboratory, TU Dresden, Dresden 01062, Germany (e-mail:	eduard.jorswieck@tu-dresden.de) }
\thanks{This research was  supported by the
	German-Israel Foundation for Scientific Research and Development under
	Grant I-1243-406.10}
}

\date{\today}
\maketitle
\begin{abstract}Design of energy efficient protocols for modern wireless systems has become an important area of research. In this paper, we propose a distributed optimization algorithm for the channel assignment problem  for  multiple interfering transceiver pairs that cannot communicate with each other. We first modify  the auction algorithm  for maximal energy efficiency and show that the problem can be solved without explicit message passing using the carrier sense multiple access (CSMA) protocol. We then develop a novel scheme by converting the channel assignment problem  into   perfect matchings on bipartite graphs.  The proposed scheme improves the  energy efficiency and  does not require any explicit message passing or a shared memory between the users.  We derive bounds on the convergence rate and show that the proposed  algorithm  converges faster than the distributed auction algorithm and achieves near-optimal performance under Rayleigh fading channels.   We also present an asymptotic  performance analysis of the fast matching algorithm for energy efficient resource allocation  and prove the optimality   for large enough number of users and number of channels.  Finally, we
provide numerical assessments that confirm the energy efficiency gains compared to the state of the art. 
\end{abstract}
\section{Introduction}
\label{section_introduction}
Communication networks have been designed to optimize conventional performance measures such as bit-error-rate, latency, and data-rate  in the past few decades. In the last few years, the issue of energy-efficient network design has gained more importance  \cite{commmag_si,jsac_si,Buzzi2015_survey_JSAC,Freng2013}. Information and communication technologies (ICT) represent about 2$\%$ of the entire world's energy consumption, and the situation is likely to reach a point where ICT equipment in large cities will require more energy than is actually available \cite{Pickavet}. For data networks, contrary to the intuition, more energy is consumed in access networks than in core networks. This happens because the number of devices in access networks (i.e. mobile terminals, base stations, and data modems installed on customers' premises) is much larger than the number of communication devices (routers, multiplexers, etc.) in the core network. This has sparked research in the field of wireless networks with a  focus on the problem of optimizing the energy efficiency (EE) at the physical layer by maximizing the ratio of a SINR-based function over the consumed energy.

Unlike cellular networks where the energy efficiency can be optimized in a centralized manner, in many cases centralized optimization is not possible. In this case distributed protocols are needed as a  practical solution to maximize the energy efficiency of the network. Several approaches had recently been suggested for optimal distributed allocation. These approaches utilize an opportunistic version of carrier sensing to determine an orthogonal allocation which is optimal or near optimal in the sense of exploiting channel diversity. The first example is the use of the well known Gale-Shapley stable marriage theorem \cite{gale1962college} to allocate spectra in a multichannel setup \cite{yaffe2010,6117764}. In \cite{6117764}, Leshem et al. demonstrated how a stable channel allocation can be obtained without any explicit transmission using carrier sensing as a mechanism to prioritize channel. Analysis of this technique for Rayleigh fading channels appeared in \cite{naparstek2012}.  In \cite{Isheden2010} and \cite{Chong2011c} a single user channel is considered and the optimization is carried out through transmit power control. In contrast, in \cite{Goodman2000}-\nocite{Power-Pricing}\nocite{Power-Receiver}\nocite{lasaulce2009introducing}\nocite{treust2010repeated} \nocite{WidelyLinear}\nocite{ZapOFDMA}\cite{belmega2011energy} and \cite{Miao2011}, multiuser interference channels are considered and a competitive scenario in which users selfishly aim at individual EE maximization is addressed. Centralized and decentralized resource allocation in multi-hop networks for energy-efficiency maximization is studied in \cite{Butt2013,Zappone2013}.

Other approaches to the distributed channel allocation problem include game theoretic bargaining solutions \cite{leshem20111,leshem20112,han2005} and distributed allocation using multichannel ALOHA \cite{cohen2013,cohen20132}.  While the stable allocation which turns out to be the greedy assignment is almost optimal for Rayleigh fading channels, it is desirable to obtain the optimal allocation. It is well known that this allocation can be computed by solving a linear programming problem \cite{kuhn1955hungarian}. However, in order to compute the optimal distributed solution \cite{naparstek2011}, \cite{naparstek2013optimal} revised the auction technique of Bertsekas \cite{bertsekas1979distributed} which requires a shared memory or price exchange between the bidders and the auctioneer. Instead of knowledge of the highest price this technique only requires knowledge of local prices. Based on the local prices, an algorithm which can be implemented using multichannel opportunistic carrier sense multiple access (CSMA) is presented and its optimality is proved.

One of the disadvantages of the distributed auction algorithm is its convergence time which might be too long in practical scenarios. It was suggested in \cite{naparstek2011} to use a simplified version of the distributed auction algorithm and look for perfect matching instead of optimal matchings to optimize energy efficiency. It was proven that the matching algorithm is asymptotically optimal for sum rate maximization and simulated results showed fast convergence time. No analysis for the expected convergence time was given in \cite{naparstek2011}. In \cite{zappone2016} energy efficiency is optimized in a distributed manner for cellular scenario using fractional programming.

In this paper, we propose a distributed optimization algorithm for the channel assignment problem  to increase the energy efficiency of multiple interfering transceiver pairs that cannot communicate with each other. We show that the problem can be solved without explicit message passing using a modified distributed auction algorithm. We then develop a novel scheme by converting the channel assignment problem of the distributed auction algorithm  into  finding perfect matchings on bipartite graphs. The proposed fast matching algorithm reduces the convergence time of the distributed auction algorithm and achieves near-optimal performance.  We analyze the performance of the fast matching algorithm for energy efficiency maximization and show that the expected energy efficiency index achieved by the matching algorithm approaches the optimal energy efficiency index for large enough number of users and number of resources. We also prove that the expected number of iterations until convergence of the fast matching algorithm is ${\cal{O}}\left(N\log(N)\right)$, where $N$ denotes the number of users. We provide numerical assessments for various wireless channels that confirm the energy efficiency gains compared to the state of the art.

The paper is organized as follows: in Section \ref{section_problem_formulation} we define the maximal energy efficiency problem. Section \ref{section_auction} discusses the distributed auction algorithm. In Section \ref{section_kabum} we present a fast converging algorithm for a relaxation to the maximal energy efficiency problem and show that the algorithm for the relaxed problem terminates within ${\cal{O}}\left(N\log(N)\right)$ iterations with high probability. The performance of the fast matching algorithm is studied in Section \ref{sect:perf}. In Section \ref{section_simulations} we discuss simulated results for the proposed algorithm. Finally, Section \ref{sect:conclude} concludes the paper.

\section{Problem formulation}
\label{section_problem_formulation}
 Before formulating the optimization problem for maximal energy efficiency, we present various definitions of energy efficiency for communication networks used in the literature.
\subsection{Definitions of Energy Efficiency}
The first and most widely used definition of EE is the ratio between the throughput and the transmit power (see \cite{Goodman2000}-\nocite{Power-Pricing}\nocite{Power-Receiver}\nocite{lasaulce2009introducing}\nocite{treust2010repeated} \nocite{WidelyLinear}\cite{ZapOFDMA}, and references therein). Another proposed metric uses the goodput in place of the throughput \cite{belmega2011energy}. In all of the above works, as far as the computation of the consumed power is concerned, only the transmit power is considered, whereas the power that is dissipated in the electronic circuitry of each terminal in order to keep the terminal active is neglected. This assumption was relaxed in \cite{Betz2008}, by defining the consumed power as the sum of the transmit power plus a constant term, independent of the transmit power, which models the circuit power needed to operate the terminal. Following \cite{Betz2008}, in \cite{Miao2011}-\nocite{Isheden2010}\cite{Chong2011c} the consumed power is also defined as the sum of the transmitted power and the circuit power. Moreover, in these papers the throughput is replaced by the achievable rate in the definition of the energy efficiency.

Essentially, each transmitter $n$ is not only interested in maximizing its own performance in terms of achieved SINR $\gamma_{n}$, but also in saving as much battery energy as possible. This trade-off is well modeled
by defining the EE of a given terminal $n$, as the ratio between the so-called efficiency function which measures the SINR-based performance of user $n$ and the power consumed to attain this performance level
\cite{Miao2011}-\nocite{Betz2008}\nocite{Isheden2010}\cite{Chong2011c}:
\begin{equation}
\mathrm{EE}_{n}=\frac{f\left(\gamma_{n}\right)}{p_{n}+P_{c,n}}.\label{eq:TUDej_EEk}
\end{equation}

In (\ref{eq:TUDej_EEk}), $P_{c,n}$ is the power that is required by the transmitter electronic circuitry to operate the device, and which is dissipated even during non-transmission periods. For further details on the circuitry power term, we refer the reader to \cite{Isheden2012} and references therein, where several power consumption models for wireless networks are developed. As for $f(\gamma)$, in principle it can be a generic increasing function of the $n$-th user's SINR, with $f(0)=0$ and such that (\ref{eq:TUDej_EEk}) tends to zero for
growing $p_{n}$. Two widely used efficiency functions are:
\begin{enumerate}
	\item $f(\gamma_{n})=R(1-e^{-\gamma_{n}})$, where $R$ is the communication
	rate and $(1-e^{-\gamma_{n}})$ is an approximation of the probability
	of correct symbol reception. A similar approximation was used
	in \cite{Power-Pricing,Power-Receiver}. Thus, $f$ is the number
	of bits that are correctly demodulated at receiver $n$ per unit of
	time.
	\item $f(\gamma_{n})=W{\rm log}(1+\gamma_{n})$, where $W$ is the communication
	bandwidth. For strictly static channels $f$ represents the $n$-th
	user's achievable rate. For quasi-static channels, the use of $f$
	for resource allocation purposes is still well-motivated in view of
	the assumption that the channel coefficients remain constant for longer
	than the resource allocation phase.
\end{enumerate}

Variations of option 1) are also available in the literature in the form of $f(\gamma_{n})=R(1-e^{-\gamma_{n}})^{M}$ and $f(\gamma_{n})=R(1-e^{-\gamma_{n}/2})^{M}$,
and in this case the function $f(\cdot)$ is an approximation of the probability of error-free reception of a data packet of $M$ symbols. An EE that considers both the case of $M>1$ and the circuit power $P_{c,n}$
was considered in \cite{Betz2008} for a single-hop system. There, it was shown that an equilibrium for the power allocation algorithm exists, but the convergence could not be proved. The techniques developed in this paper could be used to extend the results of \cite{Betz2008} to the relay-assisted scenario, as well. However, in the following we choose to focus on the equally well-motivated case of $M=1$. Thus, for any $M$, the resulting EE (\ref{eq:TUDej_EEk}) is a measure of the number of bits that are correctly decoded at the receiver, per unit of time and per Joule of energy drained from the battery of the transmitter. Moreover, all the efficiency functions that we consider result in an EE (\ref{eq:TUDej_EEk}) which is measured in bits per Joule, thus representing a natural measure of the efficiency with which each Joule of energy drained from the battery is being
used.

Two pertinent social welfare performance metrics are the average
EE and the system global EE (GEE), respectively defined by \cite{Power-Receiver}-\nocite{treust2010repeated}\nocite{lasaulce2009introducing}\cite{ZapOFDMA} as:
\begin{equation}
\label{eq_eerv}
{\rm EE}=\frac{1}{N}\sum_{n=1}^{N}\frac{f(\gamma_{n})}{P_{c,n}+p_{n}}
\end{equation}
and
\begin{equation}
\label{eq_gee}
{\rm GEE}=\frac{\sum_{n=1}^{N}f(\gamma_{n})}{\sum_{n=1}^{N}P_{c,n}+p_{n}}.
\end{equation}
Customarily, the GEE in \eqref{eq_gee} is used to describe the energy efficiency of the overall system while EE in \eqref{eq_eerv} focuses on energy efficiency of individual users.

\subsection{Problem Formulation}
We consider  $N$  transceiver pairs sharing a time slotted  frequency band divided into $K$ sub-bands. This can be seen as the open sharing model of the cognitive radio \cite{ZhaoSPM2007}. We assume $K\geq N$. This assumption can always be fulfilled: if $N>K$ than $N-K$ artificial channels with rate zero could be added and make $N=K$ \cite{bertsekas1979distributed}. Let $\mathbf{P}$ be a matrix of transmission powers where each channel is used by a single user and $P_{n,k}$ is defined as the minimal power needed by the $n$-th user to achieve a preassigned target rate $R_n$ in the $k$-th channel. We assume all the users have continuous sensing over all channels \cite{6117764}, \cite{zappone2016}. This is a reasonable assumption since the sensing power is only a small percentage of the total power in wireless networks \cite{Landsiedel2005}. We also assume that only one user can transmit on each channel in each time slot and consider alien interference from other networks as the additive noise.  Each user experiences frequency selective channel  caused  by both channel statistics and out of cell interference (since out of cell interference affects different users in a different way).

We propose a fully distributed method to maximize the energy efficiency of the system using different utility functions as described in the following:
\subsubsection{Average EE under rate constraint}
 Under the expected rate constraints,  each entry   $P_{n,k}$ of the matrix $\mathbf{P}$ is chosen to be the solution to the following ergodic rate equation
\beq
R_n=\mathbb{E}\left(\log_2\left(1+\frac{|H_{n,k}|^2P_{n,k}}{\sigma_n^2}\right)\right),
\eeq
where $H_{n,k}$ is the channel coefficient for the $n$-th user at the $k$-th frequency and $\sigma_n^2$ is the noise variance of the $n$-th receiver.
By using Jensen's inequality, we get
\beq
R_n\leq \log_2 \left( 1+\frac{\mathbb{E}\left(|H_{n,k}|^2\right)P_{n,k}}{\sigma_n^2}\right).
\eeq
The solution to the above equation gives us  to minimize the energy
\beq
P_{n,k}\geq\frac{\left(2^{R_n}-1\right)\sigma_n^2}{\mathbb{E}\left(|H_{n,k}|^2\right)}.
\label{eq_pnk}
\eeq
Using equations (\ref{eq_pnk}) and (\ref{eq_eerv}), we define the utility matrix $\textbf{U}^{\textrm{av}}$ such that ${U}^{\rm {av}}_{n,k}$ is chosen to be the optimal individual ${\rm EE^{{}}}$  for the $n$-th user in the $k$-th channel under rate constraint $R_n$
\beq
\label{eq_rateutility}
{U}^{\rm {av}}_{n,k}=\frac{R_n}{\frac{\phi\sigma_n^2}{|H_{n,k}|^2}+P_{n}},
\eeq
where $\phi=\left(2^{R_n}-1\right)$ is a function that makes the power $P_{n,k}$ used on the $k$-th channel by the $n$-th user to  satisfy a QoS requirement.  We denote $P_n$ by the minimal power of the $n$-th device for its operation including the receiver power consumption and energy consumed during idle times.

\subsubsection{Average EE under   goodput constraint}
Another problem we solve is the energy efficient channel assignment under a goodput requirement.
The achievable goodput is defined as the rate of the successfully transmitted symbols.
In this case $P_{n,k}$ is chosen to fulfill a goodput requirement:
\beq
T_n = R\left(1-{\rm SER}\right)=R\left(1-e^{-\frac{|H_{n,k}|^2P_{n,k}}{\sigma_n^2}}\right),
\eeq
for a fixed $R>0$ and $T_n$.
The solution for $P_{n,k}$ is given by
\beq
P_{n,k}=\frac{\log\left(1-\frac{T_n}{R}\right)\sigma_n^2}{|H_{n,k}|^2},
\eeq

and
\beq
\label{eq_goodutility}
U^{\textrm{good}}_{n,k}=\frac{R_n}{\frac{\phi_{}\sigma_n^2}{|H_{n,k}|^2}+P_{n}},
\eeq
where $\phi=\log\left(1-\frac{T_n}{R}\right)$.

\subsubsection{Global EE under   rate constraint}
The third problem  is the energy efficiency maximization with respect to the ${\rm GEE}$ criterion. Assuming preassigned target rates for the users, the problem simplifies into a power minimization problem under rate constraints. We assume that the instantaneous transmission power is limited per user by $P_{max}$.
For simplicity, we formulate the utility of the ${\rm GEE}$ as a maximization problem.
We define the utility matrix $\mathbf{U}^{\rm{GEE}}$ for the ${\rm GEE}$ criterion as
\beq\label{eq_utility}
U^{\rm{GEE}}_{n,k}=
\left\{\begin{matrix}
P_{\max}-P_{n,k},& P_{n,k}\leq P_{\max} \\
0 , & P_{n,k}> P_{\max}.
\end{matrix}\right.
\eeq

Combining the utility functions in \eqref{eq_rateutility}, \eqref{eq_goodutility}, and \eqref{eq_utility}, the maximum energy efficiency problem can be formulated as an integer programming problem in a general form as
\beq\label{Assign_LP12}
\bea{ll}
\displaystyle\max_{\boldsymbol\eta} \sum_{n=1}^N\sum_{k=1}^K U_{n,k}{\eta}_{n,k}\\
 s.t. \\
 \sum_k\eta_{n,k}=1,&\forall n=1,2,..,N  \\
  \sum_n\eta_{n,k}=1,&\forall k=1,2,..,K  \\
  \eta_{n,k}\in \{0,1\} ,&\forall n,k
\ena
\eeq

The constraint matrix of the problem in \eqref{Assign_LP12} is totally unimodular. Thus, the solution to the relaxed problem where we replace the integer constraint by $0\leq\eta_{n,k}\leq 1$ is also the solution to the original problem. The relaxed problem is a linear programming (LP) problem and can be solved efficiently in a centralized manner by LP solutions methods such as the Hungarian method \cite{kuhn1955hungarian}. Although the original problem in \eqref{Assign_LP12} is relaxed,  we prove that after the relaxation we can still achieve asymptotically optimal results in much lower time complexity than solving the original problem. 

In the next section, we describe a  distributed auction algorithm  for the channel assignment problem in \eqref{Assign_LP12} to achieve maximal energy efficiency of the system. Subsequently, we describe a fast matching algorithm to reduce the convergence time of the distributed auction algorithm.
\section{Distributed Auction Algorithm}
\label{section_auction}
We propose the use of a fully distributed channel assignment algorithm for maximal energy efficiency that does not require any explicit message passing or a shared memory between the users. The algorithm relies on the auction algorithm \cite{bertsekas1979distributed} and the distributed algorithm suggested in \cite{naparstek2013optimal} for sum-rate maximization. The distributed protocol consists of a bidding stage and an assignment stage. The description of the algorithm is as follows. The utility matrix $\textbf{U}$ is an $N\times K$ matrix of energy efficiency indexes and  $\textbf{C}$ is an $N\times K$ cost matrix.
 The cost of a channel $C_{n,k}$ is a unitless number that merely represents how much user $n$ wants channel $k$ in comparison to the other users. We define the profit of user $n$ from channel $k$  as the reward (i.e. energy efficiency index) minus the price of the channel $U_{n,k}-C_{n,k}$. In the initialization stage each user sets the cost for all of the channels to be $0$; i.e., $C_{n,k}=0,\forall n,k$, select $\epsilon>0$ and sets his state to unassigned. $\tilde{k}_n$ is defined as the most profitable channel of the $n$-th user:
 \beq
 \label{eq:csma}
 \tilde{k}_n=\arg\max_k\textbf{U}(n,k)-\textbf{C}(n,k).
\eeq

The distributed protocol proceeds in iterations. In each iteration two stages are sequentially performed, a bidding stage where users raise the price on their most profitable channel and an assignment stage where channels are assigned to the users who proposed the highest prices.
In the bidding stage, each unassigned user $n$ find his most profitable channel $\tilde{k}_n$ and the profits from that channel $\gamma_n$ and his second most profitable channel $\omega_{n}$
\beq
\bea{l}
\displaystyle \tilde{k}_n=\arg\max_k\left(U_{n,k}-C_{n,k}\right)\\
\displaystyle \gamma_n=U_{n,\tilde{k}_n}-C_{n,\tilde{k}_n}\\
\displaystyle \omega_{n}=\max_{k\neq \tilde{k}_n}(U_{n,k}-C_{n,k}).
\ena
\eeq
Each unassigned user raises the price on his most profitable channel by
\beq
C_{n,\tilde{k}_n}=C_{n,\tilde{k}_n}+\gamma_n-\omega_{n}+\epsilon,
\eeq
where $\epsilon$ is a predetermined positive constant that can be seen as the price of participating in the auction.
After the unassigned users update their prices, all the users bid on their most profitable channels. If user $n$ gets assigned to channel $k$ he continues to bid on that channel without raising his bid. If a user $n$ is unassigned he bids on $\tilde{k}_n$ with the new bid $C_{n,\tilde{k}_n}$. In the assignment stage each channel is assigned to the highest bidding user. A channel without bids stays unassigned and users who were not assigned to channels become unassigned. The bidding and assignment stages proceed in iterations until all the users are assigned to channels. Once all of the users are assigned to channels, no one raises his bid and as a result the assignment becomes static. When all the users are assigned we say that the algorithm has \emph{converged}.
The distributed auction algorithm appears in Table \ref{table_auction_alg}.
\begin{table}

\caption{Distributed Auction Algorithm}
\begin{tabular}{l}
\hline
Select $\epsilon>0$, set all the users as unassigned and set\\
$\textbf{C}(n,k)=0,\forall n,k$\\
\textbf{Repeat}\\
\ \ 1. Each unassigned user $n$ calculates his own maximum profit:\\
\ \ \ \ \ $\gamma_{n}=\max_k(\textbf{U}(n,k)-\textbf{C}(n,k))$\\
\ \ 2.  Each unassigned user $n$ calculates his second maximum profit: \\
\ \ \ \ \ $\tilde{k}_n=\arg\max_k(\textbf{U}(n,k)-\textbf{C}(n,k))$ \\
\ \ \ \ \ $\omega_{n}=\max_{k\neq \tilde{k}_n}(\textbf{U}(n,k)-\textbf{C}(n,k))$\\
\ \ 3. Each unassigned user $n$ updates the price of his best channel $\tilde{k}_n$ \\
\ \ \ \ \ to be $\textbf{C}(n,\tilde{k}_n)=\textbf{C}(n,\tilde{k}_n)+\gamma_{n}-\omega_{n}+\epsilon$\\
\ \ 4. All the users bid. The unassigned users bid on their new best\\
\ \ \ \ \  channel with the updated bid. The assigned users bid on the last\\
\ \ \ \ \  channel they bid on and with the same price.\\
\ \ 5. Assign channel to the highest bidder (channels with no bids \\
\ \ \ \ \ stay unassigned)\\
\textbf{Until} all users are assigned\\
\hline
\end{tabular}
\label{table_auction_alg}

\end{table}
It was proven in \cite{naparstek2013optimal} that the distributed auction algorithm for the sum-rate problem converges in finite time to a solution within $N\epsilon$ from the optimal solution.

 The distributed auction algorithm can be implemented using an opportunistic CSMA protocol without the use of explicit message passing among users. However, only coordination requirement  between users lies with an auctioneer to decide which user made the highest bid. The opportunistic CSMA can be used as an auctioneer.
 We can define the reward that each user $n$ gets from channel $k$ to be the energy efficiency of that channel $\textbf{U}(n,k)$.
Using the opportunistic CSMA scheme, each user $n$ tries to access his best profit channel as defined in  \eqref{eq:csma} with a backoff time of
$\tau_{n}=f(\tilde{k}_n)$ where $f(x)$ is a positive monotonically decreasing function. The price $\textbf{C}(n,k)$ is determined and updated if necessary as described in Table \ref{table_auction_alg}.
The prices and their corresponding waiting times must converge in a finite number of iterations as in the distributed auction algorithm.

It was shown in \cite{naparstek2013optimal} that the number of iterations needed until the convergence of the distributed auction algorithm  is bounded by ${\cal{O}}\left(N^3\right)$. In the next sections, we suggest a relaxation to the maximal energy efficiency problem. We show that the suggested relaxation is asymptotically optimal and can be solved with ${\cal{O}}\left(N\log(N)\right)$ expected number of iterations.

\section{Fast Matching Algorithm}
\label{section_kabum}
	In \eqref{Assign_LP12}, we have formulated the maximum energy efficiency channel assignment problem for arbitrary values. In the previous section, we showed that this problem can be optimally solved in a fully distributed manner using the distributed auction  algorithm. However, the expected convergence time for the near-optimal distributed auction algorithm might be too high for practical use. To speed up  the convergence time, we develop a novel scheme by considering a relaxation to the channel assignment problem where each channel can either be "good" or "bad". Essentially, channel $k$ is a good channel (corresponding to the channel gain) for the $n$-th user with respect to a properly chosen threshold. We represent a good channel by $1$ and a bad channel  by $0$. Since every channel can be either good or bad, the utility matrix of the relaxed problem becomes $\left\{0,1\right\}^{N\times K}$, which can be represented using  bipartite graphs.  Thus, the channel assignment  problem for energy efficient transmissions is reduced  finding perfect matchings on bipartite graphs.
\subsection{Maximum Cardinality Matching  on Bipartite Graphs}
To formulate the maximum energy efficiency assignment problem as a matching problem on bipartite graph and for theoretical analysis, we need  various  definitions and relevant results on  bipartite graph. These have been presented  in the Appendix A. Using these  definitions, we define the maximum cardinality matching problem as follows:  Let $G=(U,V,E)$ be bipartite graph with vertex sets $|U|=|V|=N$ and an edge set $E$. Find a matching $M$ such that $|M|$ is maximal. Here,  U,V in the bipartite graph represent the users and the channels, respectively, and the edges represent the energy efficiency of each user in each channel.

The maximum cardinality matching (MCM) problem can also be formulated as the max-energy efficiency problem (\ref{Assign_LP12}) where the reward matrix is a binary matrix with $0,1$ values. We now present an algorithm that finds a maximum cardinality matching on bipartite graphs which can be implemented in a fully distributed manner.  This iterative algorithm  assigns an unassigned user to a channel according to the following scheme: Each channel $k\in K$ is assigned a value $h_k$ that represents how many times the channel was reassigned to different users.
Let $\textbf{h}^{(i)}=\left[h_1,h_2,...,h_K\right]$ be the vector of the values of the channels on the $i$-th iteration.
At the beginning of the algorithm all the values of the channels are initialized to $0$; i.e.,
\beq
\bea{ll}
h_k^{(0)}=0,&\forall k=1,2,...,K.\nonumber
\ena
\eeq

Let $U_{\textrm{free}}$ be the set of all free users.
In each iteration, an unassigned user $u\in U_{\textrm{free}}$ is chosen and assigned to the channel with a minimal value he can access and raises its value by $1$. The MCM algorithm is summarized in Table \ref{table_alg}.
\begin{table}

\caption{Algorithm for maximal cardinality matching}
\begin{enumerate}
\item Initialize $h_v=0,\forall v\in V$, $U_{\textrm{free}}=\left\{1,2,...,N\right\}$  and set $M=\emptyset$
\item While $|M|<N$ do
\begin{enumerate}
\item Choose  $u\in U_{\textrm{free}}$
\item $j=\arg\min_{v\in n_{u}}h_v$
\item $M= M\cup (u,j)$
\item $u_{old}=\left\{u \in U:(u,j)\in M\right\}$
\item $M=M\setminus(u_{old},j)$
\item $U_{\textrm{free}}=U_{\textrm{free}}\cup u_{old}\setminus u $
\item $h_j=h_j+1$
\end{enumerate}
\item Return
\end{enumerate}
\label{table_alg}
\end{table}

\subsection{Expected Number of Iterations of Fast Matching Algorithm}
In this section, we analyze the expected number of iterations until the algorithm converges for random bipartite graphs. $G=(U,V,E)$ is called a \emph{bipartite random graph} if $G$ is a bipartite graph and the edges in $E$ are independently chosen with probability $p$; i.e.,
\beq
\bea{lll}
\Pr\left((u,v)\in E\right)=p,&\forall u\in U,&\forall v\in V.
\ena
\eeq
Denote the set of all random bipartite graphs with vertex sets $|U|=|V|=N$ and probability $p$ for an edge by $B(N,p)$.
The following known result on perfect matching in random bipartite graphs was proven by Erd\H{o}s and R\'{e}nyi in \cite{erdHos1968random} and Motwani in \cite{motwani1994average} :
\begin{other_theorems}[Erd\H{o}s and R\'{e}nyi  \cite{erdHos1968random}]
 \label{theorem_erdos}
 Let $p=\frac{(1+\epsilon)\log(N)}{N}$ and $G\in B(N,p)$ then
\beq
\lim_{N\to \infty}\Pr\left(G\textrm{ contains a perfect matching}\right)-e^{-2N^{-\epsilon}}=0.
\eeq
\end{other_theorems}

\begin{other_theorems}[Motwani \cite{motwani1994average}]\label{lemma_prob_bnp} Let $G\in B(N,p)$ where $p\geq\frac{(1+\epsilon)\log(N)}{N}$ then for every $\gamma>0$ there exists $N_{\gamma}$ such that for every $N\geq N_{\gamma}$
	\beq
	\Pr(G\in {B}(N,p))\geq 1-N^{-\gamma}.
	\eeq
\end{other_theorems}
The next theorem proven in \cite{naparstek2013random} shows that for random bipartite graphs with $p\geq\frac{(1+\epsilon)\log(N)}{N}$ the number of iterations until the convergence of the algorithm is less than $\frac{cN\log(N)}{\log(Np)}$ with high probability, where $c>0$ is a constant.
\begin{other_theorems}[Naparstek and Leshem \cite{naparstek2013random}] \label{theorem_fast}
Let $G=(U,V,E)$ be a random bipartite graph with $|U|=|V|=N$ and $p\geq\frac{(1+\epsilon)\log(N)}{N}$. Let $T$ be the number of iterations until the algorithm converges then:
\beq
\lim_{N\to \infty} \Pr\left(T\leq\frac{cN\log(N)}{\log(Np)}\right)=1.
\eeq
\end{other_theorems}
Above theorem  ensures that the fast matching finds a perfect matching with a probability that approaches $1$ in ${\cal{O}}\left(N\log(N)\right)$ iterations.

In what follows, we analyze the fast matching algorithm in wireless channels and show that the proposed scheme can be implemented in practical systems to find asymptotically optimal solution to the energy efficiency maximization problem with a small number of iterations.
\section{Performance of Fast  Matching Algorithm }\label{sect:perf}
In this section, we analyze the expected number of iterations required by the fast matching algorithm in a Rayleigh fading channel, and show that the proposed algorithm is  asymptotically optimal  for the maximally energy-efficient channel assignment problem. 

As described in the previous section, the main idea of the fast matching algorithm  is to transform $\textbf{U}$ into a binary $0,1$ matrix $\tilde{\textbf{U}}$, and then apply the matching algorithm on $\tilde{\textbf{U}}$. The transformation from $\textbf{U}$ to $\tilde{\textbf{U}}$ is done by applying a judiciously chosen  threshold $a_{n}^{\textrm{thresh}}\geq 0$ for each row:
\beq\label{eq_R_thresh}
\tilde{U}_{n,k}=
\left\{\begin{matrix}
1,& U_{n,k}\geq a_{n}^{\textrm{thresh}} \\
0 , & U_{n,k}< a_{n}^{\textrm{thresh}}.
\end{matrix}\right.
\eeq
To ensure asymptotically optimal solutions to the max-energy efficiency problem, $a_n^{\textrm{thresh}}$ must satisfy the following requirements:
\begin{enumerate}
\item Only the best channels (corresponding to the channel gain) of each user should be above $a_n^{\textrm{thresh}}$.
\item $\tilde{\textbf{U}}$ should contain a perfect matching with high probability.
\end{enumerate}
The first condition  ensures that the solution to $\tilde{\textbf{U}}$ can provide a good solution to the max-sum problem. The second condition ensures that with high probability all of the users will get assigned by the algorithm. We define a parameter $m$ such that $m\log(N)$ best channels of each user are above the threshold. 

Assume that each row $n$ of $\textbf{U}$ consists of i.i.d random variables and assume that the cumulative distribution function (CDF) of each entry of $U_{n,k}$ is given by $F_n$. A proper choice of $a_{n}^{\textrm{thresh}}$ would be
\beq
a_{n}^{\textrm{thresh}}=F^{-1}_n\left(1-\frac{m\log(N)}{N}\right),
\eeq
where $m>2$ \cite{naparstek2013random}. The choice of the threshold satisfies the first condition since only $m\log(N)$ best channels of each user are above the threshold. Theorem by Erd\H{o}s and R\'{e}nyi in \cite{erdHos1968random} ensures that with high probability $\tilde{\textbf{U}}$ contains a perfect matching if $a_{n}^{\textrm{thresh}}\geq 0$ for all $n=1,2,...,N$. Hence, the proposed algorithm can converge faster with high probability when  target rate of each user be  chosen independently such that with high probability $a_{n}^{\textrm{thresh}}\geq 0$ for all $n=1,2,...,N$. 

\subsection{Target Rates for Rayleigh Fading Channels}
\label{sec_tar_rate}
We model the channels of each user as Rayleigh fading channels; i.e., the channel attenuation $|H_{n,k}|^2$ is an exponential random variable given by
\beq
|H_{n,k}|^2=G_n\cdot F_n\cdot \frac{1}{r_n^{\alpha}},
\eeq
where $G_n$ is a global normalizing factor, $F_n$ is an exponentially distributed gain (due to the
Rayleigh fading channel with a multi-path effect), $r_n$ is the distance between the $n$-th transmitter from its receiver and  $\alpha$ is the path loss exponent.  Hence, the PDF of $|H_{n,k}|^2$ is:
\beq
f_{|H_{n,k}|^2}(x)= \lambda_ne^{-\lambda_nx},
\eeq
where
\beq
\label{eq_lambda}
 \lambda_n=\frac{r_n^{\alpha}}{G_n} =\frac{1}{\mathbb{E}(|{H}_{n,k}|^2)}.
\eeq
The instantaneous rate of user $n$ in channel $k$ is given by \cite{cover1991elements}:
\beq
R_{n,k}=\log_2\left(1+\frac{|H_{n,k}|^2P_{n,k}}{\sigma_n^2}\right).
\eeq
We now present sufficient conditions on $R_n$ such that the fast matching algorithm  converges within ${\cal{O}}\left(N\log(N)\right)$ expected number of iterations.
\begin{my_theorem}
The fast matching algorithm  converges within ${\cal{O}}\left(N\log(N)\right)$ expected number of iterations in a Rayleigh fading channel with SNR $\bar{\gamma}=\frac{P_{\max}\mathbb{E}(|{H}_{n,k}|^2)}{\sigma_n^2}$ if the following requirement is satisfied for all $n$:
\beq
\label{eq:rn_target}
R_n\leq\log_2\left(1+\bar{\gamma}\log\left(\frac{N}{m\log(N)}\right)\right)
\eeq
for the rate requirement, and
\beq
\frac{T_n}{R}\geq 1-\left(1+\frac{N}{m\log(N)}\right)^{\bar{\gamma}}
\eeq
for the goodput requirement.
\end{my_theorem}
\begin{IEEEproof}
	The proof is presented in Appendix B.
	\end{IEEEproof}

Theorem 1 provides conditions on the target rate of each user under which the proposed algorithm has a faster convergence rate. Examining \eqref{eq:rn_target} reveals that a target rate equal to the capacity of the channel satisfy the rate requirement for faster convergence. Note that  users do not  require  external knowledge or message passing among users since the conditions can be verified by each user independently.  Hence each user can choose a target rate and verify the convergence conditions without relying on other users.
\begin{figure*}[t]
	\centering
	\subfigure[Probability that the algorithm terminates after more than $N\log(N)$ iterations.. ]{\includegraphics[height=2in,width=3.5in]{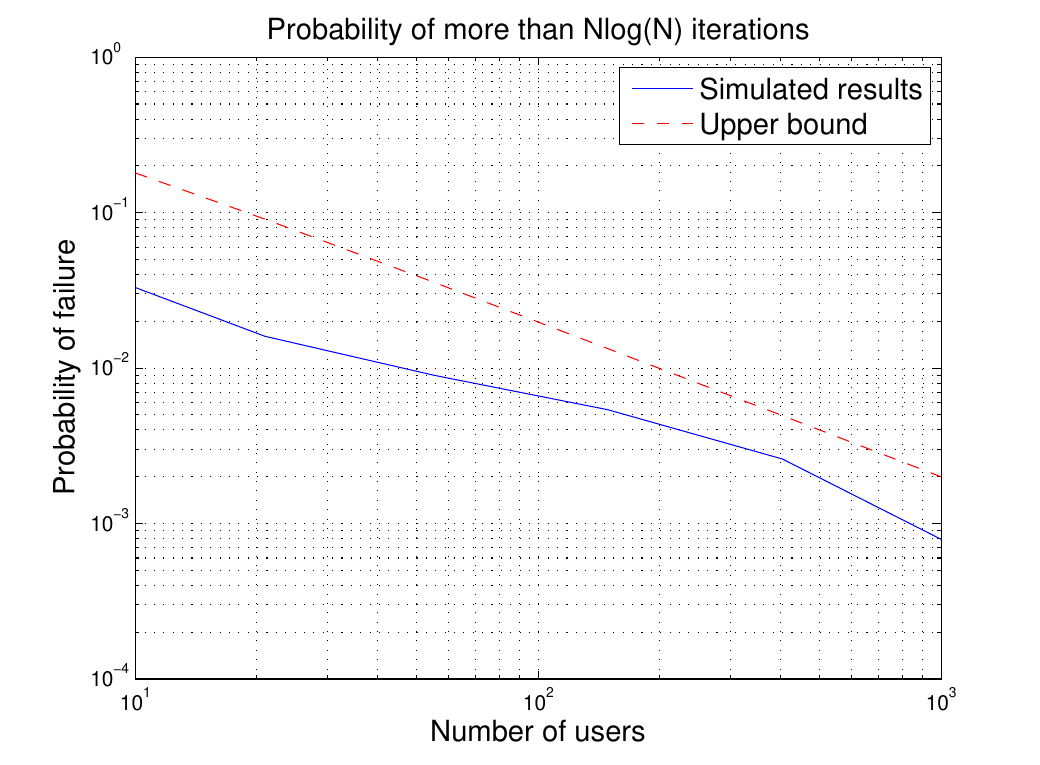}}\quad
	\vspace{-1.0mm}
	\subfigure[Expected number of iterations.]{\includegraphics[scale=0.27]{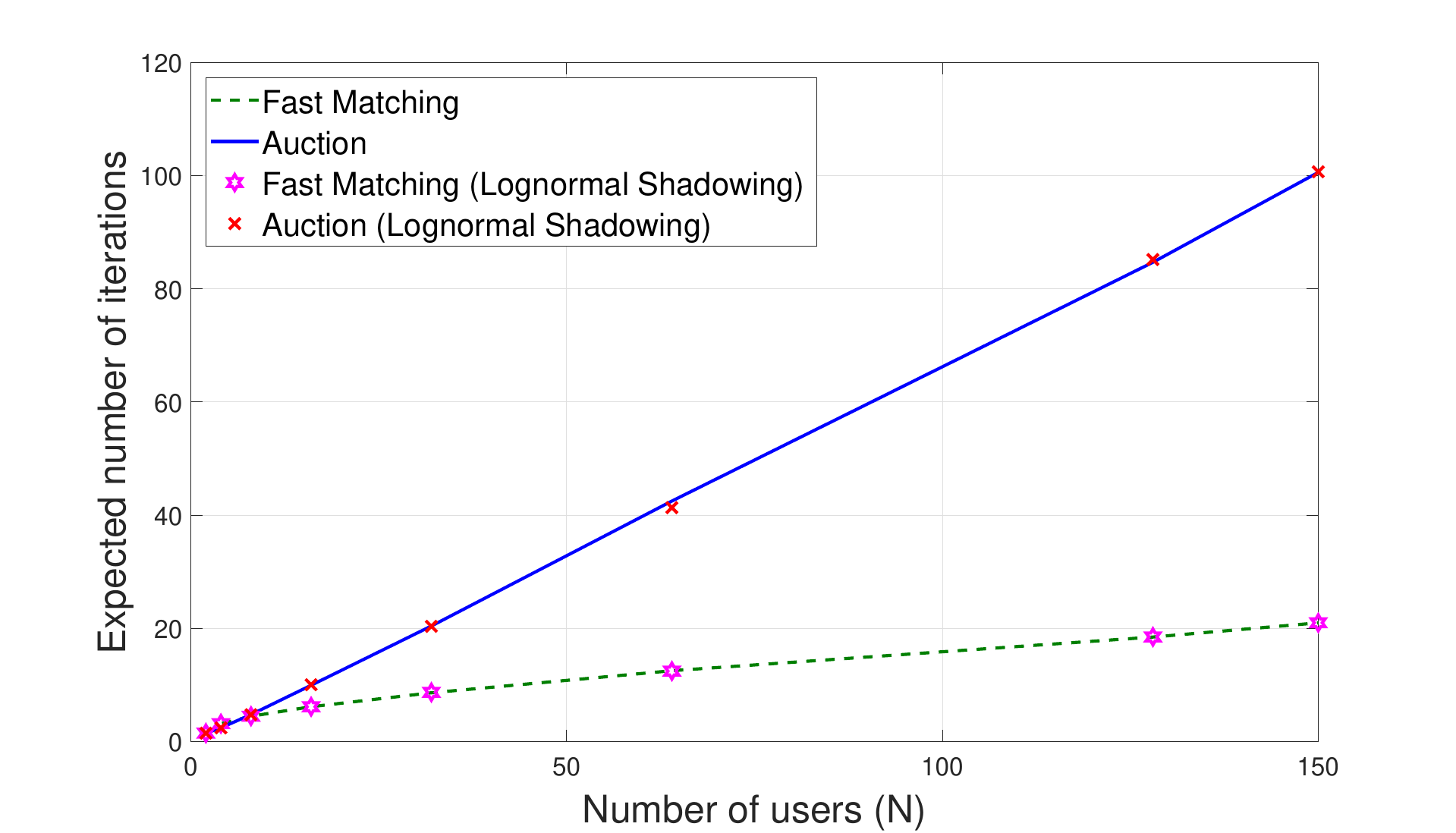}}
	\caption{Expected number of iterations required by algorithms for convergence.}
	\label{fig:iter}
\end{figure*}

\subsection{Asymptotic Optimality}
We now show that the fast channel assignment is asymptotically optimal for Rayleigh fading with properly chosen target rates. For the asymptotic analysis, we use some known results from order statistics \cite{arnold1992first}, and presents in Appendix C.

\begin{table}[t]
	\caption{Fast matching algorithm using opportunistic carrier sensing for the $n$-th user}
	\begin{tabular}{l}
		\hline
		Initialize $h_k=0,\forall k\in K$, set assigned=false\\
		set $b_n$ to be the indices of $m\log(N)$ best\\
		channels of the $n$-th user, set $N_{\textrm{iter}}=0$\\
		\textbf{Repeat}\\
		\ \ 1. If assigned=false then\\
		\ \ \ \ \ 1.1$N_{\textrm{iter}}=N_{\textrm{iter}}+1$\\
		\ \ \ \ \ 1.2 Find the channel with minimal value $\hat{j}=\arg\min_{i\in b_n}h_i$\\
		\ \ \ \ \ 1.3 Choose random backoff time $\tau_n$\\
		\ \ \ \ \ 1.4 If a busy tone was transmitted on channel $i$ before $\tau_n$ then\\
		\ \ \ \ \ \ \ \ \ 1.4.1 No transmission attempt by the $n$-th user in the current\\
		\ \ \ \ \ \ \ \ \ \ \  time slot\\
		\ \ \ \ \ \ \ \ \ 1.4.2 $h_i=h_i+1$\\
		\ \ \ \ \ 1.5 Else\\
		\ \ \ \ \ \ \ \ \ 1.5.1 Transmit a busy tone on the $\hat{j}$-th channel.\\
		\ \ \ \ \ \ \ \ \ 1.5.2 $h_{\hat{j}}=h_{\hat{j}}+1$\\
		\ \ \ \ \ \ \ \ \ 1.5.3 Set assigned=true\\
		\ \ \ \ \ 1.6 End if\\
		\ \ 2. Else\\
		\ \ \ \ \ 2.1 If a busy tone was transmitted on channel $i$ before $\tau^{\max}$ then\\
		\ \ \ \ \ \ \ set $h_i=h_i+1$\\
		\ \ \ \ \ 2.2 $N_{\textrm{iter}}=N_{\textrm{iter}}+1$\\
		\ \ \ \ \ 2.3 If $i=\hat{j}$ \\
		\ \ \ \ \ \ \ \ \ 2.3.1 No transmission attempt by the $n$-th user in the current\\
		\ \ \ \ \ \ \ \ \ \ \ time slot\\
		\ \ \ \ \ \ \ \ \ 2.3.2 Set assigned=false\\
		\ \ \ \ \ 2.4 Else transmit a busy tone on the $j$-th channel.\\
		\ \ 3. End If\\
		\textbf{Until} all users are assigned or $N_{\textrm{iter}}=(N-1)^2$\\
		If $N_{\textrm{iter}}=(N-1)^2$\\
		\ \ \ \ run the distributed auction algorithm from \cite{naparstek2013optimal}. \\
		End If\\
		\hline
	\end{tabular}
	\label{table_alg_CSMA}
\end{table}
\begin{my_theorem}
\label{theorem_asym_opt}
 Let $A_{\textrm{OPT}}^{\textrm{GEE}}$ be the optimal solution to the max-energy efficiency problem for the global energy efficiency defined in \eqref{eq_gee} and let $A_{\textrm{FMA}}^{\textrm{GEE}}$ be the solution obtained by the fast matching algorithm. If the rates satisfy \eqref{eq_R_thresh} and a perfect matching exists, then for Rayleigh fading channels:
 \beq
 \label{theorem:gee_eq}
 \lim_{N\to\infty}\frac{\mathbb{E}\{(A_{\rm{FMA}}^{\rm{GEE}}\}}{\mathbb{E}\{A_{\rm{OPT}}^{\rm{GEE}}\}}=1
 \eeq
\end{my_theorem}
\begin{IEEEproof}
	The proof is presented in Appendix D.
\end{IEEEproof}

\begin{my_theorem}\label{theorem_asym_opt2}
 Let $A^{\rm{EE}}_{\rm{OPT}}$ be the optimal solution to the max-energy efficiency problem for the individual energy efficiency defined in \eqref{eq_eerv} and let $A^{\rm{EE}}_{\rm{FMA}}$ be the solution obtained by the fast  matching algorithm. If the rates satisfy \eqref{eq_R_thresh} and a perfect matching exists, then for Rayleigh fading channels:
 \beq
 \lim_{N\to\infty}\frac{\mathbb{E}\{A^{\rm{EE}}_{\rm{FMA}}\}}{\mathbb{E}\{A^{\rm{EE}}_{\rm{OPT}}\}}=1
 \eeq
\end{my_theorem}
\begin{IEEEproof}
	The proof is presented in Appendix E.
\end{IEEEproof}

It is noted  that the asymptotic results of Theorem 2 and Theorem 3 have exactly the same meaning as any other asymptotic analysis. That is,  the performance of the fast matching algorithm  approaches the optimal solutions for large enough users and resources. Also note that in most types of asymptotic analysis, the important question: "How close is the asymptotic performance to the optimal solution?" is primarily dealt with through simulations, as discussed in the next section.

\subsection{Distributed Implementation of Fast Matching Algorithm}
We  can implement the fast matching algorithm without the use of explicit message passing using opportunistic CSMA. Opportunistic CSMA \cite{zhao2005opportunistic} is a distributed transmission protocol suggested for wireless sensor networks. Opportunistic CSMA is composed of  carrier sensing and a waiting strategy. 	Since continuous sensing of all channels by all users is assumed, each user in the network calculates a fitness measure $\psi_n$ and maps it into a waiting time $\tau_n$ based on a predetermined common decreasing function $f(\psi_n)$. Here, each user waits until the  waiting time ends and if no one transmitted on its most wanted channel then it  is allowed to transmit. Hence, the user with the highest $\psi_n$ transmits in the channel. This can be seen as a distributed winner determination algorithm where the winner gets the channel.Note that since instantaneous sensing is assumed it implies that there are no collisions. This is because the probability of two users having the same random listening time is zero.

The fully distributed fast matching algorithm for maximal energy efficiency using prioritized CSMA proceeds as follows: At the beginning of each time slot each unassigned user finds the channel with the lowest value among the  good channels (defined in ) i.e. one of the best $m\log(N)$ channels. Next, each unassigned user waits a random amount of time and if no one transmitted on any of the channels before the waiting time then that user transmits a busy signal on the channel with lowest value and raises the value of the channels by $1$. Each assigned user  waits for the maximum time allowed $\tau^{\max}$. if no user transmits on the assigned channel until $\tau^{\max}$, then the assigned user transmits on the channel.  When a user senses that a busy signal was transmitted on a channel before $\tau^{\max}$, that user raises the value of that channel by $1$.
The schematic description of algorithm is depicted in Table \ref{table_alg_CSMA}.

\begin{figure*}[t]
	
	\centering
	\subfigure[EPA channel model without log-normal shadowing. ]{\includegraphics[scale=0.25]{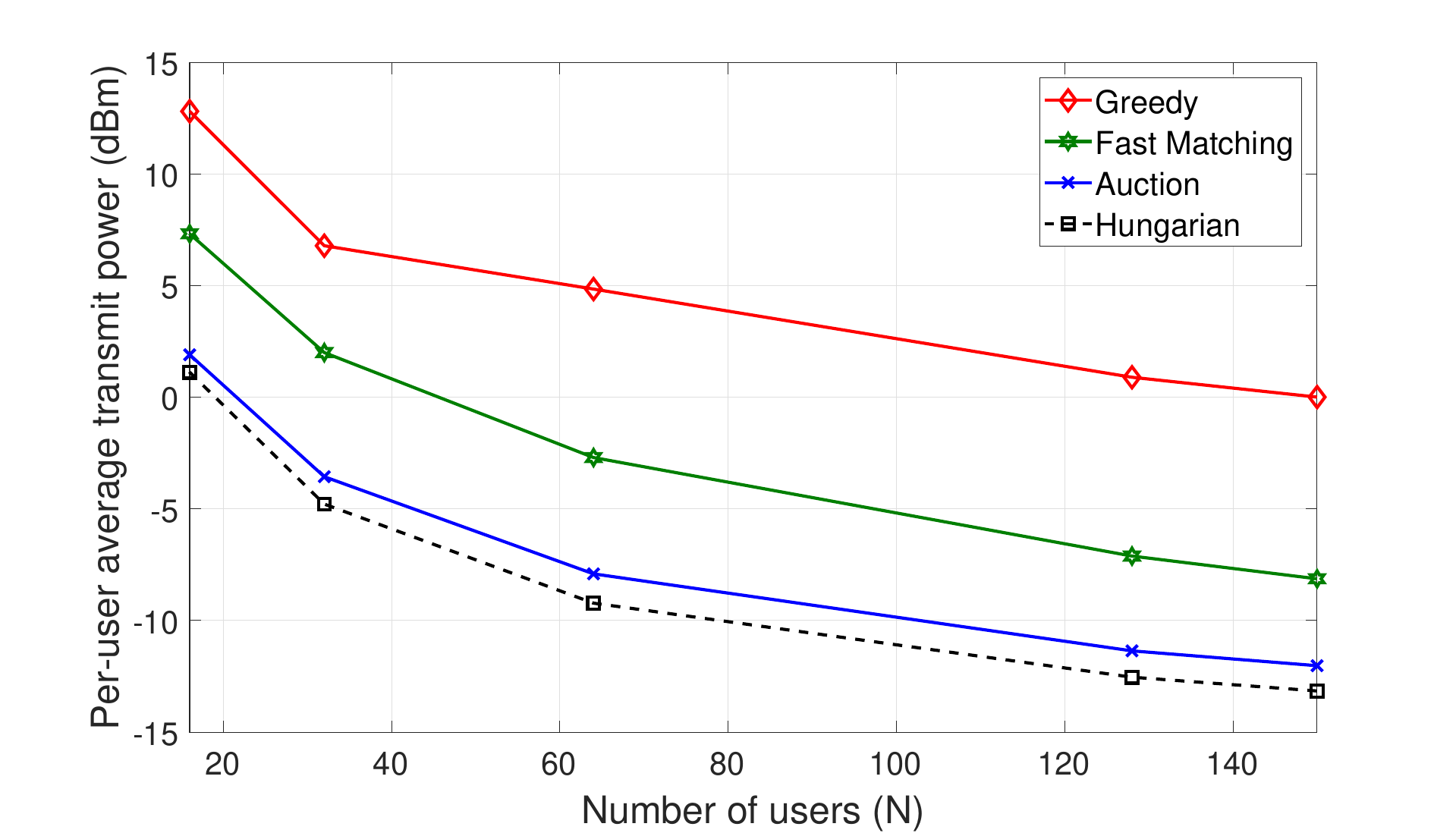}}\quad
	\vspace{-1.0mm}
	\centering
	\subfigure[EPA channel model with log-normal shadowing.]{\includegraphics[scale=0.25]{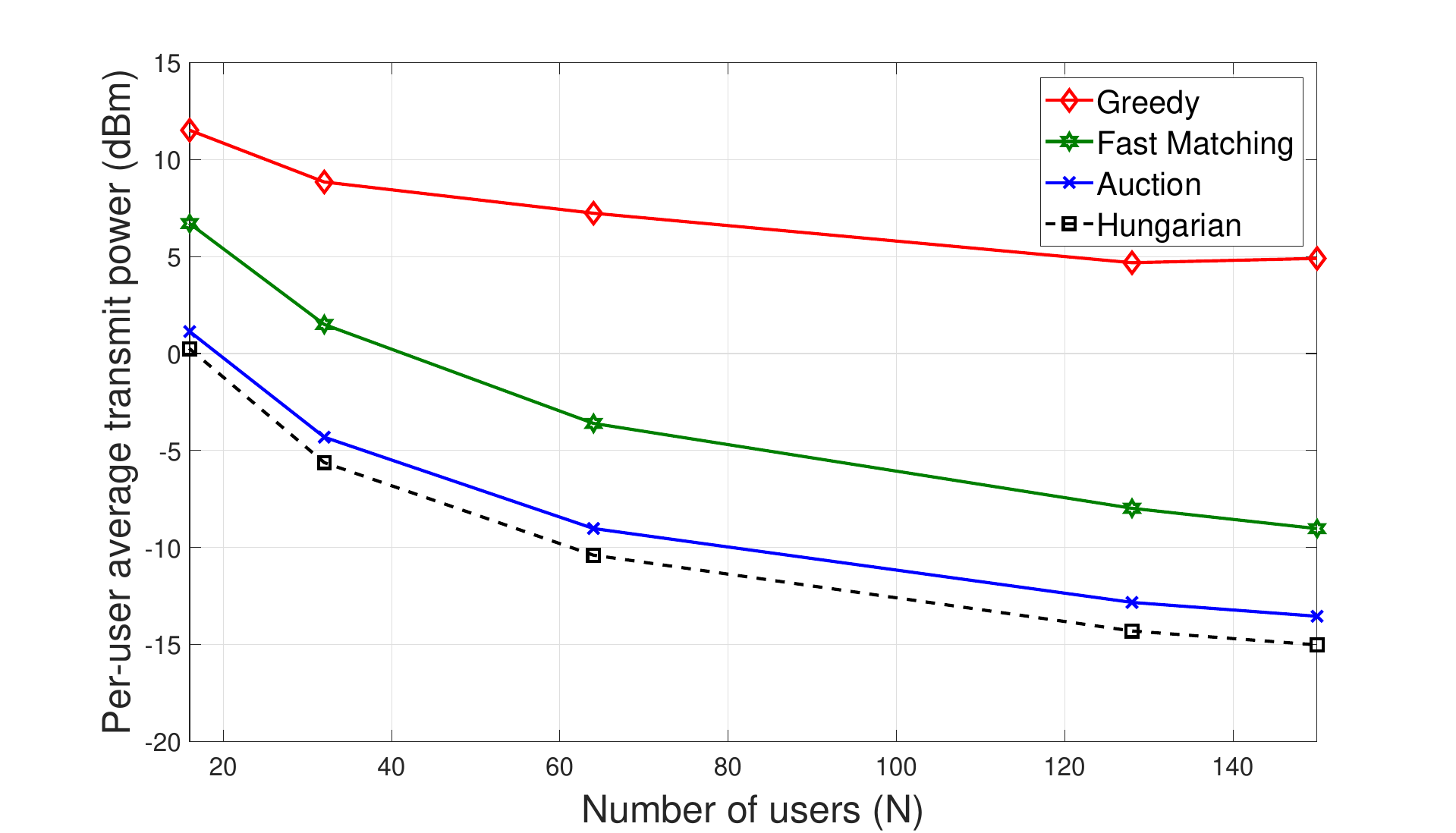}}
	\caption{Average per-user transmit power requirement to achieve a spectral efficiency of $8$ bits/sec/Hz.}
		\label{fig:power}
\end{figure*}


\section{Numerical Analysis}
\label{section_simulations}
In this section, we demonstrate the performance of the fast matching algorithm  using computer simulations.  We compared the proposed algorithm with the greedy, distributed auction, and centralized Hunganrian methods under multi-path fading and longterm shadowing effect.

 We considered a network of $N$ users  distributed uniformly in a radius of  $50$ \mbox{m} to $500$ \mbox{m} with a BS in the center. The carrier frequency was $2$ \mbox{GHz} with a per-user transmission bandwidth of $200$ \mbox{KHz}. The channel was divided into $N$ sub-channels. The channel of each user pair was generated by the extended pedestrian A model (EPA) of the LTE standard with $9$ random taps. The path loss exponent was taken as $\alpha= 3$. The users were assumed to be moving at a speed of $3$ \mbox{km/h}. We also considered channel between users to the BS to be log-normal distributed with a spreading factor of $4$ dB.  We assumed maximum spectral efficiency of $8$ bits/sec/Hz for each user. We compared the performance of the fast matching algorithm with other algorithms by computing the average power and global energy efficiency of the network. We fix the parameter  $m=2.5$ to get $m\log(N)$ channels above the threshold with $m>2$. The target rates were chosen using (\ref{eq_target_rates}) and the threshold was computed using $a_n^{\textrm{thresh}} =F^{-1}(1-\frac{m\log(N)}{N})$. The simulations were averaged over $5000$ iterations.

We investigated the expected number of iterations achieved by the fast matching algorithm compared to the distributed auction algorithm. First, we verify that the number of iterations required by the fast matching algorithm exceeds $cN\log(N)$ with a probability of less than $\frac{1}{N}$. In Fig.~\ref{fig:iter}a, we plotted the empirical probability that the fast algorithm exceeds $N\log(N)$ iterations against the theoretical bound of $\frac{1}{N}$ for $N=10...10^3$. The figure shows that the simulations support  theoretical result on the expected number of iterations for convergence. Next, we compared the expected number of iterations of the fast matching algorithm with   the distributed auction algorithm until convergence. The comparison is shown in Fig. \ref{fig:iter}b. As predicted, the expected number of iterations needed by the fast auction algorithm is smaller than the expected number of iterations  by the distributed auction algorithm.

 In Fig.~\ref{fig:power}, we demonstrate the average transmit power requirement to achieve a desired target rate (i.e. spectral  efficiency taken $8$ bits/sec/Hz) by various algorithms under two channel scenarios. It can be seen that the fast matching algorithms performs better than the  greedy method. However, the proposed algorithm requires higher transmit power than the auction method but requires significantly less number of iterations in convergence, as shown in Fig.~1b. As expected, the centralized scheme using Hungarian requires the minimum average transmit power. Moreover, the optimal distributed scheme, the auction method, performs very close to the Hungarian method.  
 
\begin{figure*}[t]

\centering
	\subfigure[Overal GEE. ]{\includegraphics[scale=0.25]{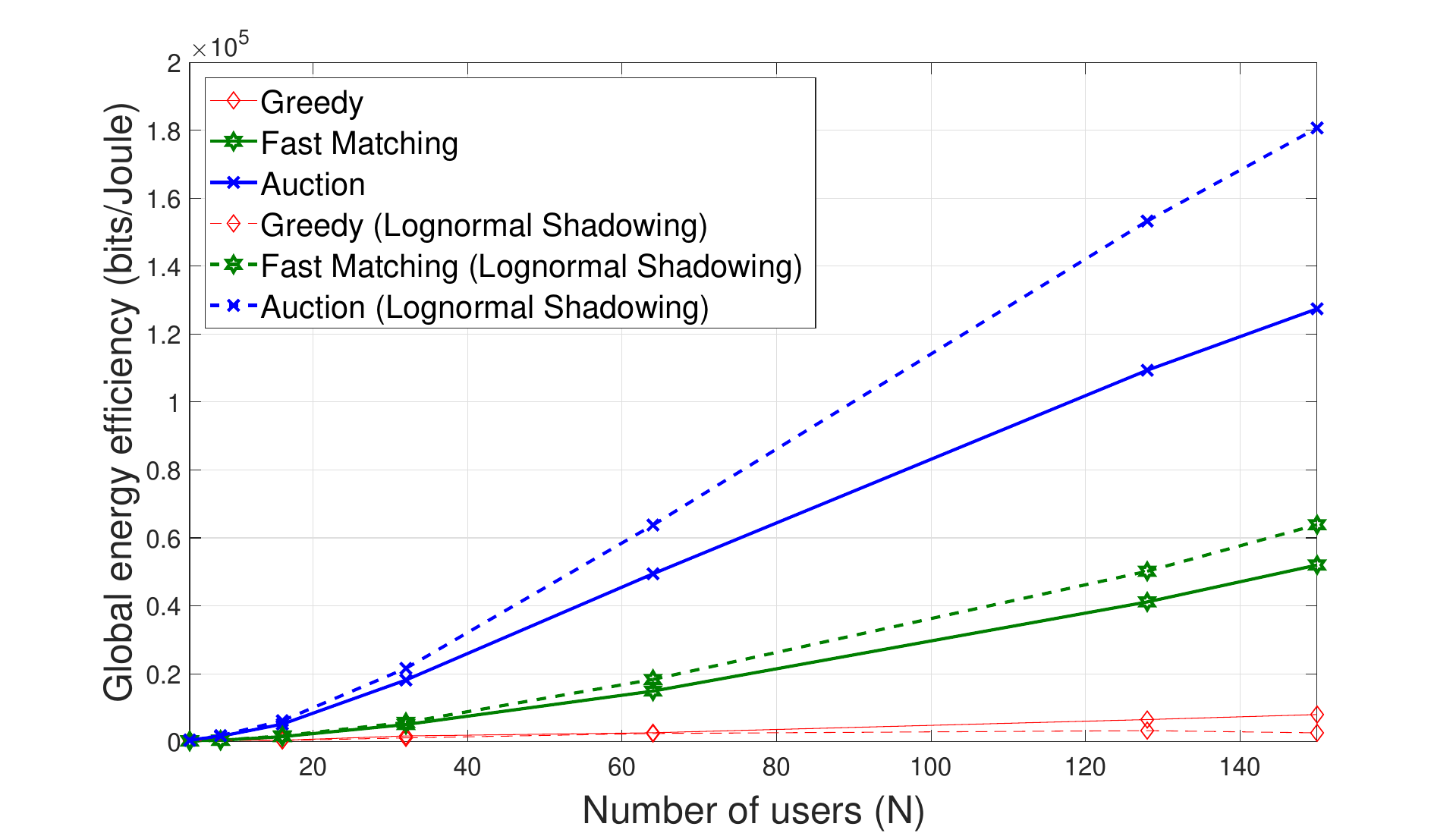}}\quad
	\vspace{-1.0mm}
		\centering
	\subfigure[GEE per iteration.]{\includegraphics[scale=0.25]{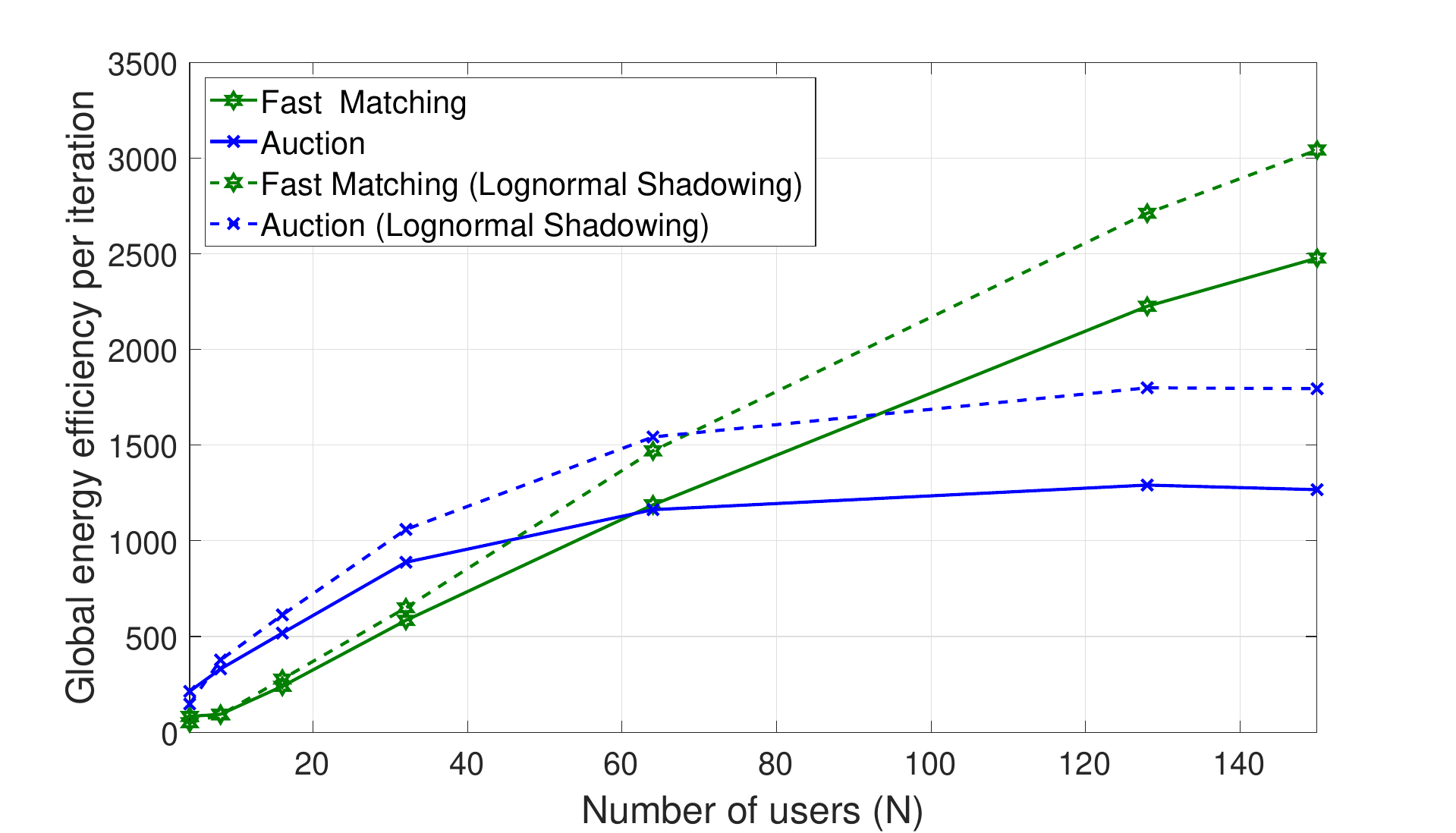}}
	\centering
		\caption{Global energy efficiency of network for the EPA channel model with and without log-normal shadowing.}
			\label{fig:gee}
\end{figure*}

We also compared the network energy efficiency of various algorithms in Fig.~\ref{fig:gee}. Although the distributed auction algorithm achieves the maximum energy efficiency (as depicted in Fig.~3a), the proposed fast matching algorithm provide greater gain per iteration (as depicted in Fig.~3b)  due to less number of iterations required for convergence.

\begin{figure}[t]
	\centering \includegraphics[scale=0.25]{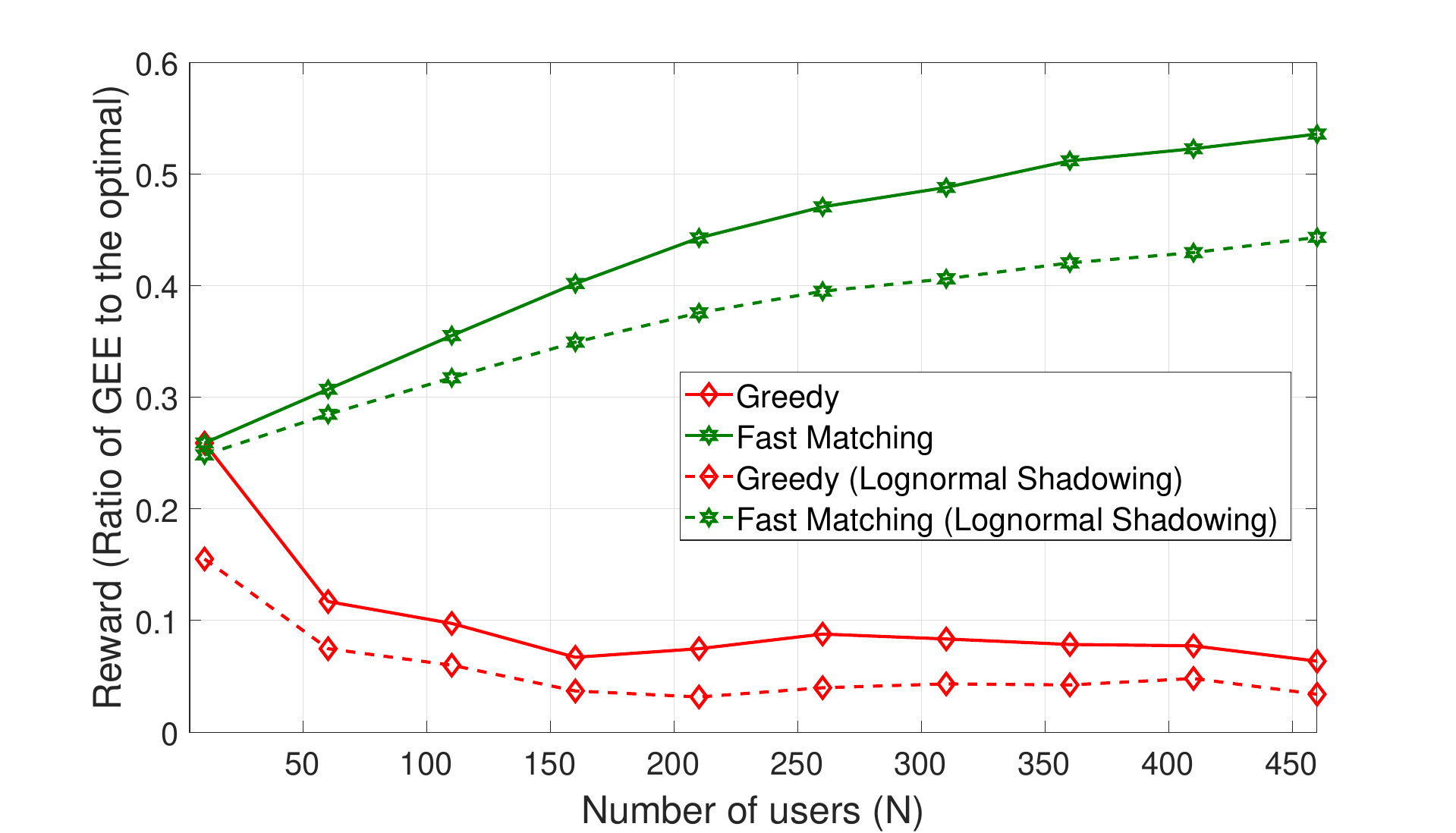}
	\caption{Asymptotic convergence of the fast matching algorithm to the optimal solution.}
	\label{fig:converge}
\end{figure}

Finally, we demonstrate the asymptotic optimality of the proposed algorithm by simulating over large network in Fig.~\ref{fig:converge}. The figure shows that the fast matching algorithm approaches the optimal auction algorithm as network size increases as proved through analysis in Theorem 2.  However, it requires much larger system sizes  to achieve the optimal solution since  the increase in the reward becomes slower as number of users increases  (i.e., when $N>200$). It can also be seen that the greedy algorithm is not asymptotic optimal and performs poorly compared with the fast matching algorithm.

\section{conclusion}\label{sect:conclude}
 We presented a fully distributed  protocol for resource allocation to optimize the energy efficiency of a wireless network.   We converted the channel assignment problem into finding perfect matchings on bipartite graph  which was shown converge within ${\cal{O}}\left(N\log(N)\right)$ expected number of iterations with high probability. The algorithm was based on a version of the auction algorithm which solves a matching problem. We also showed that under mild assumptions on the fading distribution, the fast matching algorithm  produces asymptotically optimal solutions to the  energy efficiency problem. The proposed algorithm was shown to perform better than the greedy method and achieves near-optimal performance with lesser number of iterations than the auction algorithm. The fast matching algorithm can be implemented using carrier sensing protocol to compute the channel assignments.
\section*{Acknowledgment} This research was supported by the
German-Israel Foundation for Scientific Research and Development under
Grant I-1243-406.10: Resource allocation techniques for future wireless communication networks.
\section*{Appendix A:  Bipartite Graph}
\begin{definition} Let $G=(V,E)$ be a graph with a vertex set $V$ and an edge set $E$. The \emph{neighbor set} of vertex $v\in V$ is given by
\beq
n_v=\left\{u \in V: (u,v)\in E\right\}.
\eeq
\end{definition} 
\begin{definition}  Let $G=(\hat{V},E)$ be a graph with a vertex set $\hat{V}$ and an edge set $E$. If $\hat{V}$ can be divided into two subsets $U,V$ such that
\beq
\bea{ll}
n_u\cap U=\emptyset,&\forall u\in U\nonumber\\
n_v\cap V=\emptyset,&\forall v\in V\nonumber.
\ena
\eeq
we say that $G$ is a\emph{ bipartite graph} and we denote it by $G(U,V,E)$.
\end{definition} 

\begin{definition}  Let $G=(U,V,E)$ be a bipartite graph with a vertex sets $|U|=|V|=N$ and an edge set $E$. Let $M\subseteq E$ and let $\tilde{G}=(\tilde{U},\tilde{V},M)$ be a bipartite subgraph of $G$ with vertex sets $|\tilde{U}|,|\tilde{V}|=N$ and an edge set $M$. $M$ is a \emph{matching} on $G$ if
\beq
\max_{v\in U\cup V}|n_v|=1.
\eeq
\end{definition} 
\begin{definition}  Let $G=(U,V,E)$ be a bipartite graph with vertex sets $|U|=|V|=N$ and an edge set $E$.$M$ is a \emph{perfect matching} if $M$ is a matching and $|M|=N$.
\end{definition} 
\begin{definition} Let $G=(U,V,E)$ be a bipartite graph with vertex sets $|U|=|V|=N$ and an edge set $E$. Let $M\subseteq E$ and let $\tilde{G}=(U,V,M)$ be a bipartite subgraph of $G$ with vertex sets $|U|,|V|=N$ and an edge set $M$.
A vertex $v\in U \cup V $ is \emph{free} if $|n_v|=0$ otherwise we say it is \emph{not free}.
\end{definition}

\begin{definition}   Let $G=(U,V,E)$ be bipartite graph with vertex sets $|U|=|V|=N$ and an edge set $E$. If $G$ is a random graph where each edge occurs with probability $p$ we say that $G\in B(N,p)$.
\end{definition}
\begin{definition}
	if $G\in B(N,p)$ and for any non-maximal matching $M$ there exists an augmenting path for $M$ of length at most $2L+1$ where $L=\frac{{c}\log(N)}{\log(Np)}$ and $ {c}>0$ is some constant.
\end{definition}
\begin{my_lemma} \label{lamma_T_sum_H}Let $T$ be the number of iterations until the algorithm terminates and let $h_v$ be the value of vertex $v$ at termination, then
	\beq
	T=\sum_{v=1}^N h_v
	\eeq
\end{my_lemma} 

\begin{my_lemma}
\label{lemma_neighbor}Let $G=(U,V,E)$ be a bipartite graph with vertex sets $|U|=|V|=N$ and an edge set $E$. Let $M(i)\subseteq E$ be a non maximal matching obtained by the algorithm in the $i$-th iteration. Let $h_v(i)$ be the value of vertex $v$ in the $i$'th iteration of the algorithm.
Let $D_l(i)$ be a subset of $V$ defined by:
\beq
D_l(i)=\left\{v\in V:h_v(i)\geq l\right\}.
\eeq
If $v_0\in D_l(i)$ and $(u,v_0)\in M(i)$ then
\beq
n_u\subseteq D_{l-1}(i)
\eeq
\end{my_lemma}

\begin{my_lemma}
\label{lemma_dl_1} Let $G=(U,V,E)$ be bipartite graph with vertex sets $|U|=|V|=N$ and an edge set $E$. Let $M(i)$ be a non maximal matching obtained by the algorithm in the $i$'th iteration. Let $u_0\in U$ be a free vertex such that in the $i$'th iteration of the algorithm $n_{u_0}\subseteq D_l(i)$. Let $u_1\in U$ be the end point of an alternating path $P$ with $|P|=2$ starting from $u_0$ then
\beq
\bea{ll}
v\not\in D_{l-2}(i)\setminus D_{l-1}(i)& \forall v\in n_{u_{1}}
\ena
\eeq
\end{my_lemma}
\begin{my_lemma}
\label{lemma_max_h} Let $G=(U,V,E)$ be bipartite graph with vertex sets $|U|=|V|=N$ and an edge set $E$. Let $M(i)$ be a non maximal matching obtained by the algorithm in the $i$'th iteration and let $u_0\in U$ be a free vertex such that in the $i$'th iteration of the algorithm $n_{u_0}\subseteq D_l(i)$;  then every augmenting path of $G$ on $M(i)$ starting from $u_0$ is at least of length $2l+1$.
\end{my_lemma}

\begin{other_theorems}[Berge \cite{berge1957two}]
 \label{theorem_augment}  Let $G=(U,V,E)$ be a bipartite graph with vertex sets $|U|=|V|=N$ and an edge set $E$. If $G$ contains a perfect matching there exists an augmenting path in $G$ for any partial matching $M$.
\end{other_theorems}
Using the above theorem and Lemma \ref{lemma_max_h}, we have the following results:

\begin{my_lemma}
\label{lemma_h_less_2l} Let $G=(U,V,E)$ be a bipartite graph with vertex sets $|U|=|V|=N$ and an edge set $E$. If $G$ contains a perfect matching and for any non-perfect matching $M\subseteq E$ there exists an augmenting path of length at most $2l+1$, then for every $v\in V$ at each iteration of the algorithm until the algorithm terminates
\beq
h_v(i)\leq l+1 ,\forall i,v\in V
\eeq
\end{my_lemma}
\begin{my_lemma}
 \label{lemma_worst_case}Let $G=(U,V,E)$ be a bipartite graph with vertex sets $|U|=|V|=N$ and an edge set $E$. If $G$ contains a perfect matching and  $T$ be the number of iterations until the algorithm terminates then
\beq
T\leq N(N-1)
\eeq
\end{my_lemma}

\begin{my_lemma}\label{lemma_log_path_alg} Let $G\in {B}(N,p)$ and let and let $T$ be the number of iterations until the algorithm terminates then
\beq
T\leq N(L+1).
\eeq
where $L=\frac{ {c}\log(N)}{\log(Np)}$.
\end{my_lemma}

\section*{Appendix B: Proof of Theorem 1 }
Using (\ref{eq_pnk}),  minimal power needed to achieve rate $R_n$ is given by:
\beq
{P}_{n,k}= \frac{\left(2^{R_n}-1\right)\sigma_n^2}{|{H}_{n,k}|^2}.\nonumber
\eeq
For Rayleigh fading channels, $|H_{n,k}|^2$ is exponentially distributed with the CDF:
\beq
F_{|H_{n,k}|^2}(x)=1-e^{-\lambda_n x}.
\eeq
Hence, the CDF of ${P}_{n,k}$ is given by
\beq
F_{{P}_{n,k}}(x)=e^{-\frac{\lambda_n\sigma_n^2\left(2^{R_n}-1\right)}{x}}.
\eeq
The expected number of iterations will be ${\cal{O}}\left(N\log(N)\right)$ only if there exists a perfect matching in the graph with a probability of at least $1-\frac{2}{N^{\alpha-1}}$.
From Theorem of Erd\H{o}s and R\'{e}nyi in \cite{erdHos1968random}, there exists a perfect matching with a probability of at least $1-\frac{2}{N^{m-1}}$ only
if the expected number of edges connected to each vertex is at least $m\log(N)$ for $m>2$.
Hence,  to fulfill this requirement, the expected number of channels in which each user is able to transmit without violating his power constraint is at least $m\log(N)$. A user can transmit on a channel only if the power needed on the channel to achieve the target rate is less than  $P_{\max}$. Thus,
\beq
F_{{P}_{n,k}}(P_{\max})\geq \frac{m\log(N)}{N}
\eeq
Hence, the target rates for each user must satisfy:
\beq
\label{appendix:eq1}
F^{-1}_{{P}_{n,k}}\left(\frac{m\log(N)}{N}\right)\leq P_{\max}.
\eeq
The inverse CDF of $F_{{P}(n,k)}(x)$ is given by
\beq
\label{appendix:eq2}
F^{-1}_{{P}(n,k)}(x)=\frac{\lambda_n\sigma_n^2\left(2^{R_n}-1\right)}{\log(\frac{1}{x})}.
\eeq
Using \eqref{appendix:eq1} and \eqref{appendix:eq2}, the target rates must satisfy:
\beq
\label{appendix:eq3}
F^{-1}_{{P}_{n,k}}\left(\frac{m\log(N)}{N}\right)=\frac{\lambda_n\sigma_n^2\left(2^{R_n}-1\right)}{\log\left(\frac{N}{m\log(N)}\right)}\leq P_{\max}
\eeq
Simplifying \eqref{appendix:eq3}, we get:
\beq
\label{eq_target_rates}
R_n\leq\log_2\left(1+\frac{P_{\max}\log\left(\frac{N}{m\log(N)}\right)}{\lambda_n\sigma_n^2}\right).
\eeq
Thus, the fast matching algorithm converges with an expected ${\cal{O}}\left(N\log(N)\right)$ number of iterations.

Following the same steps for the goodput requirement, we get
\beq
\label{eq_goodputs_app}
\frac{T_n}{R}\geq 1-\left(1+\frac{N}{m\log(N)}\right)^{\frac{P_{max}}{\lambda_n\sigma_n^2}}.
\eeq

Using 
$\lambda_n=\frac{1}{\mathbb{E}(|{H}_{n,k}|^2)}$, $\bar{\gamma}=\frac{P_{\max}\mathbb{E}(|{H}_{n,k}|^2)}{\sigma_n^2}$ in \eqref{eq_target_rates} and \eqref{eq_goodputs_app}, we prove the Theorem 1.
\section*{Appendix C: Order Statistics}

\begin{definition} Let $A$ be a random variable with CDF $F_A(r)$ and let $A_{1:N}<A_{2:N}<...<A_{N:N}$ be random variables obtained by taking $N$ samples from $A$ and ordering the samples in an increasing order. $A_{k:N}$ is called the $k$-th order statistics of $A$ with $N$ samples.
\end{definition}

\begin{definition} Let $k_N$ be a function of $N$ such that $k_N\to \infty$ as $N\to \infty$ and
	$\lim_{N\to\infty}\frac{k_N}{N}=0$  then $A_{N-k_N+1:N}$ and $A_{k_N:N}$ are called intermediate order statistics.
\end{definition}
\begin{definition} Let $F(x)$ be a differentiable, absolutely continuous distribution function. If
	\beq
	\lim_{x\to F^{-1}(1)} \frac{d}{dx}\left(\frac{1-F(x)}{f(x)}\right)=0
	\eeq
	then the third Von Mises condition is satisfied.
\end{definition}
\begin{other_theorems} [Falk \cite{falk1989note}]  Let $F$ be an absolutely continuous CDF satisfying one of the Von Mises conditions. Suppose $k_N\to \infty$ as $N\to \infty$ and $\lim_{N\to\infty}\frac{k_N}{N}=0$. Then there exist norming constants $\alpha_N$ and $\beta_N > 0$ such that
	\beq
	\frac{A_{N-k_N+1}-\alpha_N}{\beta_N} \xlongrightarrow{d} N\left(0,1\right).
	\eeq
	where $\alpha_N=F^{-1}\left(1-\frac{k_N}{N}\right)$ and $\beta_N=\frac{\sqrt{k_N}}{Nf\left(\alpha_N\right)}$.
\end{other_theorems}
\section*{Appendix D: Proof of Theorem 2 (GEE Asymptotic Optimality)}

We first derive probability distribution function and quantile function of $\textbf{U}(n,k)$ given that the power is lower than $P_{\max}$:
\beq
F_{{U}^{\rm{GEE}}_{n,k}}(x)=1-e^{\frac{a_n}{P_{\max}}}e^{-\frac{a_n}{P_{\max}-x}},
\eeq
where $a_n=\lambda_n\sigma_n^2\left(2^{R_n}-1\right)$
and
\beq
\label{eq:pdf_appendix}
F_{{U}^{\rm{GEE}}_{n,k}}^{-1}(\rho)=P_{\max}+\frac{a_n}{\log(1-\rho)-\frac{a_n}{P_{\max}}}.
\eeq
We now observe that the probability distribution in \eqref{eq:pdf_appendix} satisfies the third Von Mises condition resulting
\begin{align}
\begin{split}
&\lim_{N\to\infty}\mathbb{E}\left({U}^{\rm{GEE}}_{N-m\log(N)+1:N}\right)=F^{-1}(1-\frac{m\log(N)}{N}) \\&
 =P_{\max}+\frac{a_n}{\log\left(\log(N)\right)-\log(N)+\frac{a_n}{P_{\max}}}
\end{split}
\end{align}
We  now obtain simple bounds on $\mathbb{E}\left(A_{\rm{OPT}}^{\rm{GEE}}\right)$ and $\mathbb{E}\left(A_{\rm{FMA}}^{\rm{GEE}}\right)$:
\begin{align}
\begin{split}
\mathbb{E}\left(A_{\rm{OPT}}^{\rm{GEE}}\right)&\leq \sum_{n=1}^N\mathbb{E}\left(A_{N:N}\right)\\
 &=NP_{\max}-\sum_{n=1}^N\frac{a_n}{\log(N)- \frac{a_n}{P_{\max}}}
\end{split}
\end{align}
and
\begin{align}
\begin{split}
 \mathbb{E}\left(A_{\rm{FMA}}^{\rm{GEE}}\right) &\geq \sum_{n=1}^N\mathbb{E}\left(A_{N\left(1-\frac{m\log(N)}{N}\right):N}\right)\\&
=NP_{\max}-\sum_{n=1}^N\frac{a_n}{\log(N)-\log\left(m\log(N)\right)- \frac{a_n}{P_{\max}}}
\end{split}
\end{align}
It is now easy to see that
\begin{align}
\label{eq:gee:appendix}
\begin{split}
 \frac{NP_{\max}-\sum_{n=1}^N\frac{a_n}{\log(N)-\log\left(m\log(N)\right)- \frac{a_n}{P_{\max}}}}{NP_{\max}-\sum_{n=1}^N\frac{a_n}{\log(N)- \frac{a_n}{P_{\max}}}}
 \leq \frac{\mathbb{E}\left(A_{\rm{FMA}}^{\rm{GEE}}\right)}{\mathbb{E}\left(A_{\rm OPT}\right)}\leq 1.
\end{split}
\end{align}
It can be seen that
\beq
\label{eq:gee:appendix2}
\lim_{N\to\infty}\frac{NP_{\max}-\sum_{n=1}^N\frac{a_n}{\log(N)-\log\left(m\log(N)\right)- \frac{a_n}{P_{\max}}}}{NP_{\max}-\sum_{n=1}^N\frac{a_n}{\log(N)- \frac{a_n}{P_{\max}}}}=1.
\eeq
Finally, we use \eqref{eq:gee:appendix} and \eqref{eq:gee:appendix2} to get  \eqref{theorem:gee_eq} of Theorem 2.
\section*{Appendix E: Proof of Theorem 3 (EE Asymptotic Optimality)}
The proof is identical to the proof of Theorem \ref{theorem_asym_opt}. The only difference is that the CDF $F_{{U}^{\rm{av}}_{n,k}}(x)$ is given by
\beq
\label{eq:last}
F_{{U}^{\rm{av}}_{n,k}}(x)=1-e^{-\frac{\lambda_n\left(2^{R_n}-1\right)\sigma_n^2x}{R_n-c(n)x}}.
\eeq
It is easy to verify that all arguments applied to the CDF in Theorem \ref{theorem_asym_opt} can also be readily applied to the CDF in \eqref{eq:last}.

\bibliographystyle{ieeetran}
\bibliography{eeca_bibtex}

\end{document}